\documentclass[graybox]{svmult}

\usepackage{mathptmx}       
\usepackage{helvet}         
\usepackage{courier}        
\usepackage{type1cm}        
\usepackage{makeidx}         
\usepackage{graphicx}        
\usepackage{multicol}        
\usepackage[bottom]{footmisc}

\makeindex             

\begin{document}

\title*{A Survey on Approximation Mechanism Design without Money for Facility Games
\thanks{Research was partially supported by the Nature Science Foundation of China (No. 11301475) and the Nature Science Foundation
of Zhejiang Province, China (No. LQ12A01011).}}
\author{Yukun Cheng and Sanming Zhou}
\institute{Yukun Cheng \at School of Mathematics and Statistics, Zhejiang
University of Finance and Economics, Hangzhou 310018, China, \email{ykcheng@amss.ac.cn}
\and Sanming Zhou \at Department of Mathematics and Statistics, The University of Melbourne, Parkville, Victoria 3010, Australia,
 \email{smzhou@ms.unimelb.edu.au}}
%
%
\maketitle

\abstract{In a facility game one or more facilities are placed in a metric space to
serve a set of selfish agents whose addresses are their private information. In a clas-
sical facility game, each agent wants to be as close to a facility as possible, and the
cost of an agent can be defined as the distance between her location and the closest
facility. In an obnoxious facility game, each agent wants to be far away from all
facilities, and her utility is the distance from her location to the facility set. The ob-
jective of each agent is to minimize her cost or maximize her utility. An agent may
lie if, by doing so, more benefit can be obtained. We are interested in social choice
mechanisms that do not utilize payments. The game designer aims at a mechanism
that is strategy-proof, in the sense that any agent cannot benefit by misreporting
her address, or, even better, group strategy-proof, in the sense that any coalition of
agents cannot all benefit by lying. Meanwhile, it is desirable to have the mechanism
to be approximately optimal with respect to a chosen objective function. Several
models for such approximation mechanism design without money for facility games
have been proposed. In this paper we briefly review these models and related results
for both deterministic and randomized mechanisms, and meanwhile we present a
general framework for approximation mechanism design without money for facility
games.}


\section{Introduction}
\label{sec:1}

Algorithmic mechanism design \cite{NR} deals with game-theoretic versions 
of optimization problems such as task scheduling, resource allocation, facility location, 
etc. which involve one or more selfish agents who are asked to report their private information 
as part of the input. A mechanism is a function that receives the information reported by the agents, and returns an outcome possibly together with a payment scheme. An agent might lie about her information 
if doing so increases her own benefit obtained from the outcome of the game.  The goal is to design a mechanism that encourages truthfulness or strategy-proofness on the one hand and optimizes 
a related objective function on the other hand.

In mechanism design with money, the authority can use money as
compensation to the agents in order to ensure strategy-proofness.
For example, the well known Vickrey-Clarke-Groves
(VCG) mechanism \cite{N} is not only strategy-proof, but also outputs an optimal solution
to the problem of maximizing the sum of all agents' utility. However, a disadvantage \cite{R} of
this mechanism is that it must return an optimal solution for a given objective function.
Since it is often difficult to compute an optimal solution in polynomial time for many combinatorial
optimization problems, sometimes the VCG mechanism is not efficient.
Meanwhile, as pointed out by Schummer and Vohra \cite{SV2}, ``there are many important
environments where money cannot be used as a medium
of compensation, due to ethical considerations (for instance,
in political decision making) or legal considerations (for instance, in
the context of organ donations)".  Therefore, many researchers are interested in
mechanisms without monetary payment.

Procaccia and Tennecholtz \cite{PT} first proposed that approximation can be used
to obtain strategy-proofness without relying on payment and initiated a case study
in approximation mechanism design without money based on facility games. Since
then many results have been obtained by various authors on approximation mechanism
design without money for facility games. The purpose of this paper is to
give a brief review of known results in this area and meanwhile present a general
framework based on existing models. This framework will be discussed in the next
section. In Sections 3 and 4, we will give an account of results on different models for
classical facility games and obnoxious facility games, respectively. In Section 5 we discuss
possible research problems in the area.

\section{Framework}
\label{sec:2}
In any facility game there associates a set $N=\{1,2,\ldots,n\}$
 of agents, where $i$ denotes
the $i$th agent. There is also an underlying \emph{metric space} $(\Omega,d)$ whose points
are called \emph{locations}, where as usual the metric (distance function) $d:\Omega\times\Omega\rightarrow \mathbf{R}$ is
non-negative, symmetric and satisfies the triangle inequality. Each agent $i\in N$ has a
\emph{true location} $t_i\in \Omega$ that is her private information, and she reports a location $x_i\in \Omega$
that is not necessarily the same as $t_i$. We call $\mathrm{\mathbf{x}}=(x_1,x_2,\ldots,x_m)\in \Omega^n$ a \emph{location
profile}.

Assume that $k$ locations in $\Omega$ are required to be selected to put $k$ facilities. A
deterministic mechanism $f$ outputs $k$ facility locations in $\Omega$ according to a given
location profile $\mathrm{\mathbf{x}}$ without resorting to payments. In other words, $f$ is a function
$f:\Omega^n\rightarrow \Omega(k)$, where $\Omega(k)$ is the family of non-empty subsets of $\Omega$ of cardinality
at most $k$, and $f(\mathrm{\mathbf{x}})$ is the set of locations chosen for $\mathrm{\mathbf{x}}$ by $f$ where facilities will be
put.

Given a mechanism $f$ , each agent $i$ has a utility $u(f(\mathrm{\mathbf{x}}), t_i)$ whose value relies
on her true location $t_i$ and the output $f(\mathrm{\mathbf{x}})$. Since each agent is selfish, she will try
her best to maximize her utility. It may be possible for an agent to manipulate the
outcome of a mechanism to obtain more benefit by misreporting her location. Therefore,
from a game-theoretic perspective, an important goal is to design mechanisms
that are \emph{strategy-proof} (SP), in the sense that no agent can ever benefit from reporting
a false location regardless of the strategies of other agents. Sometimes we may
wish to design mechanisms that are even \emph{group strategy-proof} (GSP), in the sense
that whenever a coalition of agents lies, at least one of the members of the coalition
does not gain extra benefit from the deviation.

In approximation mechanism design, we are interested in (group) strategy-proof
mechanisms that are approximately optimal with respective to a given objective
function, where approximation is understood in the usual sense by looking at the
worst-case ratio between the optimal objective value and the value of the mechanism��s
solution to the underlying maximization problem. Based on different conditions,
such as the structure of metric spaces, the type of facilities, the number of the
facilities, etc., several models of facility games have been proposed. In the following
we summarize the major components in facility games.

\textbf{Metric Space.} So far only the following two types of metric spaces have been
considered in the literature.

\emph{Network models}: In this model a graph $G$ with each edge having a non-negative
weight is involved. We may think of $G$ as being realized as a geometric graph (in
$\mathbf{R}^3$, for example) such that the weight of each edge represents its length. The metric
space $\Omega$ is the set of points of $G$, including both vertices of $G$ and points on its
edges, and the distance $d(x,y)$ between $x\in\Omega$ and $y\in \Omega$ is the length of a shortest
path connecting $x$ and $y$ in $G$. We usually write $G$ in place of $\Omega$ in this case.

\emph{Euclidean metric space}: In this case $\Omega=\mathbf{R}^m$ for some integer $m\geq 1$ and the
distance $d$ is the usual Euclidean distance in $\mathbf{R}^m$.

\textbf{Number of facilities.} Two cases have been distinguished in the literature:

$k = 1$: In this case the unique facility provides service to all agents.

$k > 1$: In this case a set $Y$ of $k$ locations is required, and an agent is served by the
closest facility, namely, a location achieving the distance $d(Y,t_i) := \min_{y\in Y} d(y, t_i)$
between agent $i$ and $Y$.

\textbf{Type of facilities.} So far only the following two types of facilities have been
considered in the literature:

\emph{Desirable facility:} In this case all facilities (e.g. library, school, etc.) are desirable,
and each agent wants to be as close to one of the facilities as possible. As such it
is reasonable to assume that the utility $u(f(\mathrm{\mathbf{x}}),t_i)$ is a monotonically decreasing
function of $d(f(\mathrm{\mathbf{x}}),t_i)$ with only one peak. Since all facilities are desirable, we may
set $cost(f(\mathrm{\mathbf{x}}),t_i):=-u(f(\mathrm{\mathbf{x}}),t_i)$ and call it the \emph{cost function} of agent $i$. So far only
the simplest case where $u(f (\mathrm{\mathbf{x}}),t_i)=-d(f(\mathrm{\mathbf{x}}),t_i)$ for each $i$ has been studied in the
literature.

\emph{Obnoxious facility}: In this case all facilities (e.g. garbage dump, etc.) are obnoxious,
and each agent wants to be far away from all facilities. Thus the utility
$u(f(\mathrm{\mathbf{x}}),t_i)$ may be assumed as a monotonically increasing function of $d(f(\mathrm{\mathbf{x}}),t_i)$
with only one dip. The simplest case where $u(f (\mathrm{\mathbf{x}}),t_i)=d(f(\mathrm{\mathbf{x}}),t_i)$ for each $i$ has
received most attention up to now.

According to whether the facilities are desirable or obnoxious, we call a facility
game \emph{classical} or \emph{obnoxious}; each agent aims to minimize her cost or maximize her
utility, respectively.

\textbf{Strategy-proofness.} A mechanism $f$ is \emph{strategy-proof} if for any $\mathrm{\mathbf{x}}\in\Omega^n$ and
every $i$, we have $u(f(\mathrm{\mathbf{x}}),t_i)\leq u(f(\mathrm{\mathbf{x}}_{-i},t_i),t_i)$, where $(\mathrm{\mathbf{x}}_{-i},t_i)$ is obtained from $\mathrm{\mathbf{x}}$ by
replacing $x_i$ by $t_i$ but keeping all other coordinates. As a stronger requirement, $f$
is called \emph{group strategy-proof} if for any $\mathrm{\mathbf{x}}\in\Omega^n$ and $I\subseteq N$, we have $u(f(\mathrm{\mathbf{x}}),t_i)\leq u(f(\mathrm{\mathbf{x}}_{-I},t_I),t_i)$ for at least one $i\in I$, where $(\mathrm{\mathbf{x}}_{-I},t_I)$ is obtained from $\mathrm{\mathbf{x}}$ by replacing
$x_i$ by $t_i$ for every $i\in I$ but retaining all other coordinates.

\textbf{Type of mechanisms.} Two different types of mechanisms have been studied:

\emph{Deterministic mechanism:} This was discussed in the beginning of this section.

\emph{Randomized mechanism:} A randomized mechanism is a function $f: \Omega^n\rightarrow \Delta(\Omega(k))$
 where $\Delta(\Omega(k))$ is the set of probability distributions over $\Omega(k)$. In the
simplest case, the expected value $E_{Y\sim f}[d(Y,t_i)]$ may be defined as the cost or the
utility of agent $i$ in classical facility games or obnoxious facility games, respectively.

\textbf{Objective function.} The decision maker (or the mechanism designer) is interested
in (group) strategy-proof mechanisms that also do well with respect to optimizing
a given objective function.

Similar to the $k$-median and $k$-center problems \cite{DH,KH1,KH2}, for classical facility
games researchers have so far considered minimizing the \emph{social cost} $SC(f,\mathrm{\mathbf{x}}):=\sum_{i=1}^n cost(f(\mathrm{\mathbf{x}}),t_i)$ or the \emph{maximum cost} $MC(f;\mathrm{\mathbf{x}}):= \max_{i=1,\ldots,n}cost(f(\mathrm{\mathbf{x}}),t_i)$.
Similar to the $k$-maxisum and $k$-maximin problems \cite{C,T1,Z}, for obnoxious facility
games researchers have considered maximizing the \emph{obnoxious social welfare}
$SW(f;\mathrm{\mathbf{x}}):=\sum_{i=1}^n u(f(\mathrm{\mathbf{x}}),t_i)$ or the \emph{minimum utility}
 $MU(f;\mathrm{\mathbf{x}}):=\min_{i=1,\ldots,n}u(f(\mathrm{\mathbf{x}}),t_i)$.

In summary, a facility game consists of: a set $N$ of $n$ agents; a metric space
$(\Omega,d)$ which may be continuous or discrete; a subset $\{t_1,\ldots,t_n\}$
of $\Omega$, $t_i$ being the true location of agent $i$; a set of $k$ facilities to be installed at $k$ (not necessarily
distinct) locations in $\Omega$; a utility function $u:\Omega(k)\times\Omega\rightarrow \mathbf{R}$ taking non-negative
values which usually relies on $d(f(\mathrm{\mathbf{x}}),t)$, where $x=(x_1,\ldots,x_n)\in\Omega^n$, $t\in\Omega$, and
$f:\Omega^n\rightarrow\Omega(k)$ is a deterministic mechanism; and a non-negative objective function
$F:\Omega(k)\times\Omega^n\rightarrow \mathrm{\mathbf{R}}$ to be maximized, which is usually defined in terms of $u(f(\mathrm{\mathbf{x}}),t_i)$,
$1\leq i\leq n$. We are interested in designing a mechanism $f$ that is strategy-proof or
even group strategy-proof on the one hand, and on the other hand outputs a good
solution for any location profile in the sense that the approximation ratio
\begin{eqnarray*}
\sup_{x\in\Omega^n}\frac{\max_{Y\in \Omega(k)}F(Y,\mathrm{\mathbf{x}})}{F(f\mathrm{\mathbf{(x}}),\mathrm{\mathbf{x}})}
\end{eqnarray*}
is as small as possible. Different specification of the components above gives rise
to different models for approximation (deterministic) mechanism design without
money for facility games.

In approximation randomized mechanism design, the distance function, the utility
function and the objective function are all random, and we can give a similar
framework by considering the expected values of the corresponding random variables.

\section{Classical Facility Games}
\label{sec:3}

\subsection{Single facility games}

In the case $k=1$, the preferences are
\emph{single peaked} in the sense that the outcome is less preferred by each agent when it is further from her ideal locations.
Beginning with \cite{M1}, single peaked preferences and their
extensions have been extensively studied in the social choice
literature. In this subsection, we summarize known results on finding a facility
location in different metric spaces that minimizes the social cost
or the maximum cost.

If the objective is to minimize the social cost, Procaccia and Tennecholtz \cite{PT}
proposed a GSP optimal mechanism which returns the location of the median agent as the
facility location when all agents are located on a path. This mechanism is GSP since
an agent can manipulate the output only by misreporting her location to be
on the opposite side of the median. Moreover, the median also
minimizes the social cost,
because for any location with distance
$\epsilon>0$ to the median, at most $\lfloor n/2\rfloor$ agents are within distance
$\epsilon$ to the facility and all other agents are away from the facility by
at least $\epsilon$. Similarly, if the graph is a tree, Alon \emph{et al}. \cite{AFPT} gave a
mechanism that outputs the median of the tree as the facility's location.
Such a mechanism is also an optimal GSP mechanism.

When all agents are located on a graph $G$ containing a cycle $C$, Schummer and Vohra
\cite{SV1} showed that if a deterministic mechanism $f:G^n\rightarrow G$ is an SP mechanism that
is onto $G$, then there is a cycle dictator, that is, there exists $i\in N$ such that for all
$\mathrm{\mathbf{x}}\in C^n$, $f(\mathrm{\mathbf{x}})=x_i$. Based on such a characterization,
Alon \emph{et al}. \cite{AFPT} obtained a tight SP lower bound of $n-1$ on the approximation ratio
for any graph $G$ that contains a cycle. For the randomized version, they designed a mechanism which returns
a facility location $x_i$, $i\in N$, with probability $1/n$. This mechanism is SP
with approximation ratio $2-(2/n)$ for any
general graph. They showed further that such a
mechanism is GSP if and only if the maximum degree of the graph is two.

If the objective is to minimize the maximum cost, the problem of designing an SP
mechanism is simpler compared with deterministic mechanisms. Since Schummer
and Vohra \cite{SV1} showed that strategy-proofness can only be obtained by dictatorship,
Alon \emph{et al}. \cite{AFPT} considered the mechanism given by $f(\mathrm{\mathbf{x}})=x_1$ for all $\mathrm{\mathbf{x}}\in G^n$,
that is, agent 1 is a dictator. It can be proved that such a mechanism is a GSP
2-approximation mechanism. On the other hand, Procaccia and Tennecholtz \cite{PT}
showed that a deterministic SP mechanism cannot achieve an approximation ratio
better than 2 even if the underlying graph $G$ is a path. Thus the dictatorship mechanism of
Alon \emph{et al}. \cite{AFPT} has the best approximation ratio with respect to the maximum cost.
Procaccia and Tennenholtz \cite{PT} proved that a randomized SP mechanism has
approximation ratio at least $3/2$ on a path. They also gave
a matching GSP upper bound of $3/2$ by using the \emph{Left-Right-Middle} (LRM) Mechanism, which, for a given $\mathrm{\mathbf{x}}\in G^n$, chooses $\min_{i}x_i$ and $\max_{i}x_i$ with probability $1/4$ respectively, and chooses the midpoint of the interval $[\min_{i}x_i,\max_{i}x_i]$ with probability $1/2$.
When the agents are on a circle, Alon
\emph{et al}. \cite{AFPT} proposed a randomized SP mechanism with approximation ratio $3/2$ that
combines two mechanisms: the LRM mechanism if the agents are located on
one semicircle, and the Random Center Mechanism otherwise.
When $G$ is a tree, they showed that
there is a randomized SP $(2-\frac2{n+2})$-approximation mechanism that,
for a given $\mathrm{\mathbf{x}}\in G^n$, outputs
$x_i$ for each $i\in N$ with probability $1/(n+2)$ and the center of the
tree with probability $2/(n+2)$.
They further proved that $2-O\left(\frac{1}{2^{\sqrt{\log n}}}\right)$ is a lower bound
on the approximation ratio for any SP randomized mechanism.

Procaccia and Tennecholtz \cite{PT} considered a natural extension of the classical
single facility games, in which one facility should be located but each agent controls
multiple locations. As before, the objective is to minimize the social cost
or the maximum cost. However, the cost of an agent now depends on the objective
function. If the objective is to minimize the social cost, the cost of agent $i$ is
defined as $cost(y,\mathrm{\mathbf{x}}_i) = \sum_{j=1}^{w_i}d(y,x_{ij})$, where $y$ is the location of the facility and
$\mathrm{\mathbf{x}}_i=(x_{i1},\ldots,x_{i w_i})$ is the location set controlled by agent $i$. If the objective is to minimize the maximum cost, the cost of
agent $i$ is $cost(y,\mathrm{\mathbf{x}_i}) = \max_{j=1,\ldots,w_i}d(y,x_{ij})$. For the social cost, they directly applied
the deterministic mechanism by Dekel \emph{et al}. \cite{DFP} that returns the median $med(\mathrm{\mathbf{x}}')$
of $\mathrm{\mathbf{x}}'=(med(\mathrm{\mathbf{x}_1}),\ldots,med(\mathrm{\mathbf{x}}_n))$.
Dekel \emph{et al}. \cite{DFP} also showed that this mechanism
is a GSP 3-approximation mechanism and provided a matching lower bound. Furthermore,
Procaccia and Tennecholtz \cite{PT} designed a simple randomized mechanism
to return $med(\mathrm{\mathbf{x}}_i)$ with probability $\frac{w_i}{\sum_{j\in N}w_j}$. This mechanism is SP, and when
$n = 2$ its approximation ratio is $2+\frac{|w_1-w_2|}{w_1+w_2}$. Subsequently, Lu \emph{et al}. \cite{LWZ} extended
the result about the approximation ratio to $3-\frac{2\min_{j\in N}w_j}{\sum_{j\in N}w_j}$
for any $n$ and obtained the
lower bound $1.33$ by solving a related linear programming problem. For the maximum
cost, they proposed a GSP 2-approximation deterministic mechanism and a
$(3/2)$-approximation randomized mechanism. Since the multiple location
setting is the same as the simple setting stated before when $w_i= 1$, $i\in N$, any
lower bound for the simple setting holds here as well.

\subsection{2-facility games}

When the objective function is the social cost and the network is a path, Procaccia
and Tennecholtz \cite{PT} showed that the mechanism that outputs an optimal solution
for a given $\mathrm{\mathbf{x}}\in G^n$ is not strategy-proof. They gave the following GSP $(n-1)$-
approximation mechanism: choose the leftmost and the rightmost points, and
constructed an instance to show that $3/2$ is a lower bound on the approximation
ratio for any SP deterministic mechanism. Later,
 Lu \emph{et al}. \cite{LWZ} improved such
lower bound to 2, designed a randomized $n/2$-approximation mechanism, and explored a lower bound of 1.045 for randomized mechanisms.
 Moreover, Lu \emph{et al}.
\cite{LSWZ} proved that the $(n-1)$-approximation deterministic mechanism given in \cite{PT}
is asymptotically optimal. They constructed an instance on a path and explored the
lower bound $(n-1)/2$ on the approximation ratio by employing two key concepts:
partial group strategy-proofness and image set. In the case when all agents are on a
circle, they designed a GSP deterministic mechanism with an $(n-1)$-approximation
ratio which asymptotically matches the lower bound $(n-1)/2$. Lu \emph{et al}. \cite{LSWZ} also
obtained an SP 4-approximation randomized mechanism called the \emph{Proportional
Mechanism}: the first facility is allocated uniformly over all reported locations; the
second facility is assigned to another reported location with probability proportional
to its distance to the first facility.

If the objective is to minimize the maximum cost, only Procaccia and Tennecholtz
\cite{PT} contributed some positive results in the case when all agents are on a
path. For the deterministic version, they applied the same deterministic mechanism as the one for the social cost model. By exploring the
characterization of the structure of the optimal solution, they proved that the approximation ratio of such a mechanism is 2, and
provided a matching SP lower bound. Furthermore, they designed a randomized SP
$5/3$-approximation mechanism. Compared with the deterministic case, the randomized
mechanism for this model is much more complicated and the authors applied
some new ideas: randomizing over two equal intervals, unbalanced weights at the
edges, and correlation between the two facilities. These strategies play a crucial
role in satisfying the delicate strategy-proof constraints and break the deterministic
lower bound of 2. The lower bound of any randomized SP mechanism is proved to
be $3/2$.

\subsection{k-facility games with $k\geq 3$}

For $k$-facility games with $k\geq3$, most known results focus on the objective of
minimizing the social cost. McSherry and Talwar \cite{MT} first used differentially private
algorithms as almost strategy-proof approximate mechanisms. The main advantage of such
an algorithm is that it can control any agent's influence on the outcome so that
any agent has limited motivation to lie.
McSherry and Talwar
presented a general differentially private
mechanism that approximates the optimal social cost within an additive logarithmic
term. Unfortunately, the running time of this general mechanism is randomized exponential-time.
Subsequently, Gupta \emph{et al}. \cite{GLMRT} presented a computationally efficient differentially private
algorithm for several combinatorial optimization problems. Based on \cite{MT}, Nissim \emph{et al}. \cite{NST} considered \emph{imposing
mechanisms} which can penalize liars by restricting the set of allowable post-actions
for the agents. They combined the differentially private mechanisms of \cite{MT} with an
imposing mechanism and obtained a randomized imposing SP mechanism
with a running time in $k$ for $k$-facility location. The mechanism approximates the optimal average social
cost, namely the optimal social cost divided by $n$, within an additive term of roughly
$1/n^{\frac13}$.

In contrast to \cite{NST}, Fotakis and Tzamos \cite{FT} tried to design an SP mechanism
with standard multiplicative notion of approximation. They considered the \emph{winner-imposing}
mechanism which chooses $k$ reported locations of agents to build facilities. If an agent's reported
location is chosen to put a facility, then she is served by this facility and her service cost
is the distance between this facility and her true location. If an agent's reported
location is not chosen, then she is served by a facility closest to her true
location. Thus the winner-imposing mechanism can penalize an agent without
money only if she succeeds in gaining more benefit in the mechanism. Fotakis and Tzamos
proved that the winner-imposing version of the Proportional Mechanism in \cite{LSWZ} is
an SP $4k$-approximation randomized mechanism. Moreover, they addressed the \emph{facility
location game} in which there is a uniform facility opening cost, instead of a
fixed number of facilities. The authority should place some facilities so as to minimize
the social cost and the total facility opening cost. For this game, they showed
that the winner-imposing version of Meyerson's randomized algorithm in \cite{M2} is an
SP 8-approximation mechanism. Meanwhile, they presented a deterministic nonimposing
GSP $O(\log n)$-approximation mechanism when all agents are on a path.
In addition, Escoffier \emph{et al}. \cite{EGTPS} considered a facility game to locate $n-1$ facilities
to $n$ agents. They studied such a game in the general metric space and trees for the
social cost and the minimum cost, and provided lower and upper bounds on the
approximation ratio of deterministic and randomized SP mechanisms.

\section{Obnoxious Facility Games}

For obnoxious facility games on a path, the preferences are known as \emph{single-dipped},
meaning that the worse allocation for each agent is the one that places the
facility right by their home, and that locations become better as they are further
away. In the past a few years, a lot of work \cite{BBM,HD,IN,M3,PS} was focused on
characterizations of the strategy-proofness for the single-dipped preference. Cheng
\emph{et al}. \cite{CYZ} initially studied approximation design without money for obnoxious facility
games with the objective of maximizing the obnoxious social welfare. In this
section, we survey some known results in this domain.

Cheng \emph{et al}. first proposed group strategy-proof mechanisms to locate one facility
with respect to different network topologies. In particular, if all agents are on a path,
they viewed such a path as an interval with left endpoint $a$ and right endpoint $b$.
Since this model is related to the literature on approximation algorithms for the 1-
maxian problem \cite{CG,T1,T2,Z} from an algorithmic perspective, it is well known that
one of the two endpoints must be an optimal facility location for $\mathrm{\mathbf{x}}\in G^n$. Thus they
regarded the two endpoints as the candidates for the facility locations and designed
a GSP 3-approximation deterministic mechanism, which outputs $a$ if the number
$n_2$ of agents on the right-hand side of the interval is larger than the number $n_1$ of
agents on the left-hand side, and $b$ otherwise. By a similar idea, they presented
two GSP deterministic mechanisms respectively when all agents are on a tree or a
circle, and proved that the approximation ratio of each mechanism is 3. Later, Han
\emph{et al}. [12] provided the matching strategy-proof lower bounds for each model on
different networks. Furthermore, when all agents are on an interval, Cheng \emph{et al}. \cite{CYZ}
also gave a randomized mechanism which returns $a$ and $b$ with probability $\alpha$ and
$1-\alpha$, respectively, where $\alpha=\frac{2n_1n_2n_2^2}{n_1^2+n_2^2+4n_1n_2}$.
 They proved that such a randomized
mechanism is GSP and has achievable approximation ratio $3/2$. When all agents
are on a general network, a GPS 4-approximation deterministic mechanism and
a trivial GSP 2-approximation randomized mechanism were derived. In addition,
the deterministic mechanism was shown to be asymptotically optimal by using the
characterization of strategy-proofness for general networks \cite{IN}.

Recently, Cheng \emph{et al}. \cite{CHYZ} considered a new model of obnoxious facility games
that has a bounded service range. In this model each facility can only serve the
agents within its service range due to the limited service ability. Each agent wants
to be far away from the facilities. On the other hand, she must stay within at least
one facility's range, otherwise she can not receive any service. Cheng \emph{et al}. first
studied the case when all agents are on an interval, which is normalized as $[0,1]$,
and the service radius is some $r$ with $1/2\leq r\leq 1$. Compared with the previous
model without service range, this new model is more complicated since more than
one facilities may be needed and it is no longer true that one of the endpoints must
be an optimal solution. According to the value of $r$, Cheng \emph{et al}. selected different
candidates for the facility locations. To be specific, if $3/4 \leq r \leq 1$, points $r$ or $1-r$
are designated as the facility locations; otherwise, they locate one facility at $1/2$
or two facilities at $0$ and $1$ respectively. Thus they designed a GSP deterministic
mechanism and a GSP randomized mechanism. When $1/2 \leq r <3/4$ or $3/4 \leq
r \leq 1$, the approximation ratio of their deterministic mechanism is $8r-1$ or $\frac{2r+1}{2r-1}$,
respectively, and the approximation ratio of their randomized mechanism is $4r$ or
$\frac{2r}{2r-1}$, respectively. Meanwhile, they also proved a lower bound for any strategy-proof
deterministic mechanism by constructing different instances, which is equal to $4r-1$
if $1/2 \leq r < 3/4$, $1/(2r-1)$ if $3/4\leq r < 5/6$ and $3r-1$ if $5/6 \leq r < 1$.

\section{Conclusion}

We reviewed some known results on approximation mechanism design
without money for facility games. By comparing our general framework in Section 2 and
what we surveyed in Sections 3 and 4, it should be clear that a lot of interesting problems
remain open and different models may be considered by specifying the components
in the framework. For example, one may investigate various cases where the space of locations
is more involved, such as a multi-dimensional Euclidean space or a specific network
other than paths, trees and cycles.

For obnoxious facility games, except the results in \cite{HD} there are no other results
in the case when the objective is to maximize the minimum utility. Han and Du
proved that there is no any SP deterministic mechanism with finite approximation
ratio for this objective. We believe that results can be obtained by
using the differentially private algorithm mentioned in Section 3.3 to design almost SP
mechanisms.

For facility games with a limited service ability, the only known result is about the
obnoxious facility game on interval $[0,1]$ with a service range $1/2 \leq  r \leq 1$. It would
be interesting to investigate classical and obnoxious facility games with different
types of restrictions to service ability in different metric spaces. In particular, one
may consider the obnoxious facility game on $[0,1]$ when $0 < r < 1/2$. It seems
challenging to find a general SP mechanism corresponding to the value of $r$.

Finally, closing the gap between the lower and upper bounds on the approximation
ratios of deterministic or randomized mechanisms for some models is also a
significant research problem.

%
%
%

\end{document}